# Fundamentals of the Orbital Conception of Elementary Particles and of their Application to the Neutron and Nuclear Structure

G. Sardin (*)

**Abstract**

An alternative approach to the Standard Model is outlined, being motivated by the increasing theoretical and experimental difficulties encountered by this model, which furthermore fails to be unitary. In particular, the conceptual uneasiness generated by the excessive multiplicity of fundamental elements of the Quark Model, 36 different quarks whose cohesion needs 8 different types of gluons, has logically led some physicists to propose a variety of quark substructures in an effort to reach unity. However, these hazardous attempts will without any doubt guide particle physics to fall into an abyss, in view of the already too highly dubious content of QCD.

In order to avoid the forward escape corresponding to the attribution of a substructure to quarks and to stand away from the conceptual strangling to which the Standard model has led, we have instead opted for different fundamentals. These, in contrast to those of the Standard Model, are extremely simple and based on the assumption of a single fundamental corpuscle, of dual manifestation as corpuscle and anticorpuscle, to which is always associated an orbital that determines the structure of particles. In such a frame particles differentiate through the diversity of quantum states of their structuring orbital, in contrast to the strategy used by the Standard Model based instead on the particle's multiplicity of composition through the variety of the quark's content, furthermore limited to hadrons. Instead the orbital conception of particles is unitary, unifying all of them as well as their interactions. As an outstanding feature, nuclear forces derive from the neutron orbital structure, based on a proton core and a shell. This shell constitutes the cohesive element of nuclear structure.

**Résumé**

Une alternative au Modèle Standard est esquissée, motivée par les difficultés croissantes, théoriques et expérimentales, que rencontre ce modèle qui par ailleurs n'arrive pas à être unitaire. En particulier, le désarroi conceptuel provoqué par l'excessive multiplicité d'éléments fondamentaux du Modèle des Quarks, 36 differents quarks dont la cohesion requiert 8 different types de gluons, a de façon logique conduit certains physiciens à proposer une variété de sous-structures des quarks. Cependant ces essais hasardeux conduiront sans aucun doute la physique des particules à un précipice, compte tenu du déjà haut contenu douteux de la QCD.

Afin d'éviter la fuite en avant correspondant à l'attribution d'une sous-structure aux quarks et de se maintenir en marge à l'étranglement conceptuel auquel le Modèle Standard a conduit, nous avons opté pour des fondements différents. Ceux-ci, contrairement à ceux du Modèle Standard, sont extrêmement simples et se basent sur l'hypothèse d'un unique corpuscule fondamental, de nature duale existant comme corpuscule et anticorpuscule et auquel est toujours attribué une orbitale qui détermine la structure des particules. Dans ce context les particules se différencient par la diversité d'états quantiques de leur orbitale structurelle, contrastant avec la stratégie qu'utilise le Modèle Standard qui se base sur la multiplicité de composition des particules obtenue grâce à la diversité de leur contenu en quarks différents, à la fois limitée aux seuls hadrons. De façon aventageuse la conception orbitale des particules est unitaire, s'appliquant à toutes elles et unifiant aussi leurs interactions. Comme conséquence remarquable, les forces nucléaires découlent de la structure orbitale du neutron, basée en un noyau formé par un proton et une enveloppe, laquelle constitue l'élément cohésif de la structure nucléaire.

# I. INTRODUCTION

## I.1. Fundamentals of the orbital conception of elementary particles and nuclides

The basic concepts on which stands an alternative and unitary description of objects at the Fermi scale ($10^{-15}$ m) are introduced in terms of conceptual physics (1,2). This includes elementary particles and their interactions (3-8,12,19), as well as the building blocks formed by two of them, the proton and the neutron, which lead to the large set of stable and unstable nuclides (9-16).

The fundamentals are self-sustained and converge into a unitary conception of all objects at the mentioned scale and of their interactions, named Quantum Orbital Structure (QOS). The concepts developed are not derived from theoretical developments but straightforwardly from the respective experimental data (17,18), however they end up concurring with quantum field theory, more specifically with the fundamentals of Quantum Electro-Dynamics (QED).

Similarly to QED, within QOS vacuum is considered to be populated of virtual quanta, but the QOS goes one step further by attributing them a specific structure defined by a pair of structural corpuscular carriers of opposite unitary charges spinning together into a common orbital ($c^+, c^-$) defining the structure of these neutral virtual quanta. Furthermore, they are considered to be virtual photon-like quanta having similarly an intrinsic speed. Their ground state is considered to be aenergetic, however any energy transfer brings them into energetic states represented by neutral elementary particles, such as the photon, neutrinos, neutral pions, etc. according to the quantum state acquired.

Still, these neutral quanta may brake into their two structural components, leading to two separate charged quanta of opposite sign and being represented by the diverse charged particles and antiparticles depending on the diversity of quantum state of their structure orbital. The details of the axiomatic base of the QOS have been edited elsewhere (1).

## I.2. Duality, intrinsic celerity, intrinsic confinement and spin

In order to outline the conceptual frame developed, some of the primordial characteristics of Nature in its material dimension will be first commented on. One of them is duality, which appears as a standard, such as the one from negative and positive unitary electric charge or particle and antiparticle, and it will be retained as the primordial base of the fundamentals developed.

Extending hence the precept of duality to the structure of elementary particles, they will be considered to derive from a dual fundamental system, which can dissociate into its two parts. Another basic characteristic stands on self-celerity, such as with the photon. Still another relevant one rises from self-confinement, which combined to self-celerity generates the structuring orbital, spin, magnetic moment, interactions, stability, etc.

## I.3. Massive and massless energy

Another fundamental concept stands in the differentiation between massive energy and massless energy. In the present frame this differentiation derives directly from the basis of the fundamental system which applies to all elementary particles. The massive or massless nature of elementary particles is defined by the residue between two antagonist energy components of the fundamental system representative of any elementary particle. According to the predominant component the particle has a mass and no intrinsic celerity or is massless or near massless and consequently acquires an intrinsic celerity.

The best-known representative of this latter case corresponds to the photon, the massless carrier of the electromagnetic field, when in its free state. In the virtual state, the photon self-celerity provided by its massless state allows it to act also as interacting force carrier of the electric and magnetic fields.

## I.4. Elementary particles: the electron, proton and neutron

All elementary particles are considered to derive from a single fundamental system and to correspond to different manifestations rising from the system's different quantum states. To illustrate the fundamental concepts used these will be applied to three most representative massive elementary particles, the electron, the proton and the neutron. The restrictive selection of these three massive particles is motivated by the immediate application of their orbital conception to the nucleus structure and cohesion, providing a novel interpretation of nuclear forces.

The model can easily be extended to all elementary particles and applies to the massive as well as to the massless particles and to their interactions of short and long range. Among all elementary particles, the electron and the proton constitute very special particles, being the sole two stable massive particles. Furthermore, in the present conceptual frame they appear as two differentiated but closely related forms of a single elemental system and may be looked at as constituting a fundamental tandem of twin particles which present a deep mutual affinity.

The neutron, which is unstable in its free state, is considered made of the unstable union between a positively charged, dense, heavy, and stable core and a negatively charged, light, unstable shell, a union that builds up its composed structure. In fact the neutron as such does not exist in any bonded state through strong interaction since this irremediably implies sharing its shell and thus the loss of its free identity.

### I.5. The atomic nucleus and the nature, saturation and short range of nuclear forces

The orbital axiomatic concepts applied to the proton and the neutron will be extended to the atomic nucleus, considering at first a few light nuclei such as the deuteron $_1H^2$, the triton $_1H^3$, the helions $_2He^3$ and $_2He^4$, and afterwards the whole of nuclides. One of the most immediate applications of the orbital theory of elementary particles is concerned with nucleons and specifically with the saturation of nuclear forces and the limitation of isotone, isobar and isotope families. The orbital model offers a novel and straightforward explanation to the nature, saturation, and short range of nuclear forces.

### I.6. Link with quarks and partons

Within the orbital context quarks as well as partons must be reinterpreted, at best (19-23). They can no longer keep their status of particles. The orbital theory may incorporate specific aspects of the parton and the quarks, but at the cost of attributing them a different nature. From the orbital perspective, partons can no longer stand to be point-like objects forming a cloud behaving as a quasi-ideal gas and leading, e.g. to the proton. Partons can only be retaken as representing the point-like density of presence whose distribution would define the body orbital of elementary particles.

Also, in an attempt to integrate the quark model (which is not unitary, applying only to hadrons) within the orbital model (which is unitary, applying to all elementary particles), quarks could be regarded as a way to typify the quantum substates of the hadrons' structure and gluons as a way to typify internal and external interactions. In other words, from the orbital outlook quarks could express the subdivision (flavors and colors) of the main orbital structure of particles into an orbital substructure.

Gluons could be regarded as expressing the complex way in which these substructures would be interrelated. In any case this reinterpretation effort of the quark model does not allow regarding quarks and gluons as particles but only as sub-structural virtual entities. Whatever, quarks cannot preserve their status of fundamental material bodies in the orbital context, from which all the diverse elementary particles are conceived as generated by the different allowed orbital quantum states of a unique fundamental corpuscle.

Let us recall that quarks, if seen as particles, are quite strange ones; they form three double families: u-d, c-s and t-b and since they possess three colors they are eighteen. With the antiquarks they form a set of thirty six. To this already excessively large set of fundamental elements should be added another set of fundamental elements constituted by eight types of gluons, being in charge of the bonding and confinement of quarks. A total of forty four particles as elemental building blocks without applying to all elementary particles, cannot be seen as an efficient reductive system.

Furthermore, besides their color charge which is quite mysterious and their perturbing non integer electric charge of 1/3 and 2/3, quarks cannot be directly detected but only deduced, so they may only reach the status of virtual particles. This high content of strangeness on the part of the fundamentals of the quark model produces a conceptual uneasiness due to the feeling that it may be far away from an ascertained and unitary conception of elementary particles.

We think that the Standard Model approach to the nature of elementary particles (hadrons), based on a diversity of composition (quarks) instead of a diversity of quantum states of a unique structure, corresponds to a misconception. Recall that there are only four stable particles: two massive ones, i.e., the electron and the proton, and two massless ones (or near massless), i.e., the photon and the neutrino. All other particles (misleadingly so called) are unstable and together with resonances have extremely short lifetimes, except the neutron, whose lifetime is comparatively quite long. This sole fact already strongly suggests the conception of them as excited states of a unique basic structure.

Furthermore, the Standard Model is unbalanced and incomplete, being hypertrophied with respect to hadrons, which apart from the proton and neutron, are extremely short-lived and foreign to the building of matter, hence only of marginal interest. In counterpart the Standard Model has nothing to propose for the structure of such crucial particles as the photon, electron and neutrinos, which are stable and ubiquitous.

So, let us propose a conceptual alternative which appeals instead to a unique fundamental element, essentially characterized by the attribution of an intrinsic, mutable orbital and a dual-form corpuscle-anticorpuscle. Such a rudimentary conceptual tool reaches nevertheless to give account for all elementary particles, for their four types of interaction, for the nuclear structure and furthermore converges into a unitary conception of matter. Let us briefly make explicit the axioms of the orbital theory and the corresponding description of archetypical systems, from elementary particles, such as the nucleons, to atomic nuclei.

## II. FUNDAMENTALS

The core axioms of the orbital conception of elementary particles stands in their being structured by the orbital of an elemental corpuscle. From a first basic approach let us introduce the main features of the elemental corpuscle and of its associated orbital.

### II.1. The elemental corpuscle

The three fundamental concepts mentioned in the introduction, namely duality, self-celeration and self-confinement, are now applied to this corpuscle and form the most basic axioms.

**a.** The corpuscle is assumed to be elemental and unique but dual, i.e. it has no substructure but exists as corpuscle and anticorpuscle, with opposite unitary electric charge.

**b.** The corpuscle is a self-celerated object, i.e. it has an intrinsic celerity just like the photon. In fact, the photon celerity derives from the corpuscle self-celerity (celerity is here used equivalently to speed, however celerity is considered more appropriate when it is constant and speed when it is variable).

**c.** The corpuscle is self-confined within a closed space in which it describes an orbital. In contrast to the photon, which has a lineal celerity, the corpuscle path is self-curved with a radius of femtometric size ($10^{-15}$ m or a Fermi), sticking to a weak equilibrium between induced centripetal and centrifugal forces. The corpuscle cannot avoid having an associated orbital, so it always manifests itself dressed with a body orbital which shields its detection.

### II.2. Elementary particles

Elementary particles are considered to be the diverse manifestations of the corpuscle orbital. The diversity of elementary particles corresponds to the diversity of structures and quantum states of the corpuscle orbital. Let us specify now the orbital axioms associated with elementary particles.

**a.** Singly charged elementary particles are formed by a single corpuscle to which is associated an orbital wave function $\psi$ which conjugated product ($\psi\psi^* = \psi^2 = \Psi$) defines the particle orbital structure (from now on the structure orbital of particles will be designed by $\Psi$ instead of $\psi^2$).

**b.** Neutral elementary particles are two-component particles, composed of a corpuscle and an anticorpuscle, spinning together and leading to two superposed orbitals ($\Psi^+$ and $\Psi^-$), which may be identical or not ($\Psi^+ \equiv \Psi^-$ or $\Psi^+ \neq \Psi^-$).

**c.** The quantum states of the corpuscle orbital define the elementary particle and any change in the orbital quantum states leads to a different elementary particle.

**d.** The stability of the elementary particle is fixed by the stability of its structure orbital.

**e.** The particle's spin and the magnetic moment both derive directly from their orbital structure.

**f.** The net energy of the orbital structure defines the intrinsic energy of the elementary particle. The orbital net energy is fixed by the balance between a massless energy which derives from the spinning of the corpuscle electric charge which has hence an electromagnetic nature, and a massive energy which derives from an antagonistic restoring force. The prevailing energy component determines the manifested nature of the particle energy. Hence the particle may be massive, such as e.g. the proton, as well as massless, such as the photon.

**g.** The equilibrium of the particle orbital structure is determined by two antagonistic forces $F_1$ and $F_2$. One is centripetal (a Lorentz like force) and the other one is centrifugal (restoring like force). These two forces are seen as action and reaction.

The force $F_1$ is centripetal and derives from the spinning of the corpuscle electric charge, which nature is thus electromagnetic. Its magnitude expressed from the c.g.s system is expressed as:

$$F_1 = (e^2/m_0 * c^2) * (1/r^2) * (p * c) \quad (1)$$

where $e$ and $m_0$ are the electron electric charge and mass, $c$ is the celerity of light, $r$ is the orbital radius and $p$ is the corpuscle momentum.

The force $F_2$ is an antagonistic restoring force which is thus centrifugal and has a massive nature. Its magnitude is expressed as:

$$F_2 = (1/r) * (p * c) \quad (2)$$

The two forces reach equality ($F_1 = F_2$) for $r = r_0 = e^2/m_0 c^2$, which is the electron classical radius. If the two forces are not equal then the net force $\Delta F$ induces a variation $\Delta r$ of the orbital radius ($\Delta r = r_0 - r$). The orbital acquires hence a net energy:

$$E = E_1 - E_2 = (F_1 - F_2) * \Delta r = \Delta F * \Delta r \quad (3)$$

**h.** Short range interactions, weak and strong, derive directly from the particle's orbital structure defined by its wave function $\Psi$. The strength of the interaction is determined by the degree of overlapping (extension and density) of the orbital structures and also by their quantum states.

**i.** Long range interactions derive from the interchange of massless particles in a virtual state, i.e. of dual orbital systems composed of a corpuscle and an anticorpuscle in a massless orbital state. This state allows the particle to express its energy into celerity instead of mass, enabling it to act as long range carrier.

## III. ARCHETYPICAL ORBITAL SYSTEMS

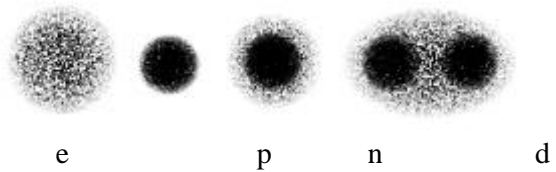

   e         p         n         d

*Fig. III.1: Orbital structure of the electron, proton, neutron and deuteron*

### III.1. Proton and Electron

Within the orbital context, the proton (1) is considered to be constituted by a single orbital spun by a corpuscle with positive charge and defined by its structure wave function $\Psi_p^+$. It is thus shaped by a unique charge distribution, (Fig.1) whose size is of the order of the Fermi. The proton structuring orbital has a relatively high net energy of 938.27 MeV, which expressed in mass corresponds to 1.0072765 amu. The high energy of the proton shaping orbital makes it usually appear and behave as a dense and massive hard body. The proton spin and magnetic moment are both considered to derive from its structuring orbital and as a matter of fact to constitute significant tracks of it.

Its apparent size is observation-dependent, i.e. according to the type of collision it may appear point-like or as a body with finite size. When the impinging particles are electrons of very high energy with a corresponding wavelength much shorter than the Fermi, in crossing over the proton structuring orbital they perceive it as a voluminous body, due to their high resolution power. Instead, electrons of low energy with wavelength much larger than the Fermi are unable to perceive the corpuscle spatial distribution which defines the particle structuring orbital, and due to an insufficient resolution power they coarsely perceive the proton as a point-like object.

The electron (1) is instead structured by an orbital (Fig.1), defined by its structure wave-function $\Psi_e^-$ and with an energy of only 0.51 MeV, equivalent to a mass of 0.55 x $10^{-3}$ amu. The formulation of the total rest energy of the electron is: E = T + V, where T expresses its structure dynamical energy and V expresses that any charged particle has a potential energy with respect to its neutral original system. The potential energy V derives from the dissociation energy of the initial dual system ($c^+, c^-$) from which proceeds the electron ($c^-$).

The electron rest energy of 0.51 MeV, considered to be a potential energy with respect to the original dual neutral system, it implies thus that the net value of its structural energy component T must be null for E = T + V to be equal to 0.51 MeV. This is the reason that makes the electron structural orbital undetected and considered instead to be point-like, a feature which proceeds from its structuring corpuscle. The spin and magnetic moment of the electron arise from the specific characteristics of its structuring orbital.

The spinning dynamics of the orbital is thus totally converted into magnetic moment, without any massive component, in accordance with the fact that the structure's inner net kinetic energy is null (T=0). Hence, the electron mass rises exclusively from the potential energy (V) proceeding from its previous dissociation from its tandem opposite charge. Further developments on the proton and electron, such as their unification and quantization, have been exposed elsewhere (1).

**III.2. Neutron**

In terms of structuring orbitals, the neutron is considered to be a dual particle, composed of two orbitals (Fig.1), of opposite electric charge (1) defined by a composite structural wave-function $\Psi_n = \Psi_p^+ + \Psi_s^-$. These orbitals are quite different. One is highly energetic, of smaller size and plays the role of a positively charged core ($\Psi_p^+$), while the other one is much lighter, slightly wider and acts as a negatively charged shell ($\Psi_s^-$). The core orbital is in fact considered to correspond to the proton structuring orbital with a mass of 938.27 MeV/$c^2$, while the wrapping orbital mass is equal to only 1.29 MeV/$c^2$ since the neutron mass is equal to 939.56 MeV/$c^2$. Both neutron spin and magnetic moment correspond to the resultant components of the two orbitals.

The detection of the neutron shell is made difficult by its low energy with respect to the core energy. The high-energy collision of an impinging electron with a neutron corresponds in fact to a double collision, firstly with the neutron shell of 1.29 MeV and secondly with its core of 938.27 MeV. According to the collision conditions the first collision may or may not be significant. Furthermore, depending on the shell spatial distribution it may leave bareheaded part of the neutron core, ending up to behave as a loose and elusive shield. As a consequence, high energy collisions of massive particles with neutrons are not sensitive to the neutron shell and they essentially hit with its core. The collision with the core proton is brought back to the previous considerations about the proton.

The associative superposition of the two structuring orbitals, core ($\Psi_p^+$) and shell ($\Psi_s^-$), of the neutron is unstable. The spontaneous disintegration of the free neutron corresponds to the exclusive disintegration of its shell ($\Psi_s^-$). The products of the neutron degradation provide indeed a perfect fingerprint of its structure, i.e. the core corresponding to a proton; the shell orbital that restructures when getting free into an electron orbital; and a neutrino that carries away part of the energy released by the transition of the neutron wrapping orbital into the electron orbital.

The emission of a neutrino during the orbital transition allows also to preserve the spin conservation. Relative to the neutron stability, one important point which derives from the orbital conception of the neutron stands in the excess energy of the wrapping orbital with respect to the mass of the two resultant massive particles of its disintegration, i.e. the proton and the electron. This excess of energy induces the decay of the neutron since it consequently slips to a lower total mass. The total energy and the spin are preserved through the emission of a neutrino. Besides, the neutron may be seen as an elemental nucleus, the most elemental one since its wrapping orbital is formed by a single carrier and contains a single proton core.

The disintegration of the neutron obeys the weak interaction. In terms of the orbital model this corresponds to the sole loss of the neutron shell ($\Psi_s^-$). In getting free, the shell restructures into a new orbital corresponding to the electron ($\Psi_e^-$). Since the neutron shell has an energy of 1.29 MeV, i.e. 0.78 MeV higher than the electron energy of 0.51 MeV, this energy excess is dissipated into kinetic energy transmitted in part to the electron and in part into the emission of a neutrino of variable energy. One of the neutron shell's most relevant properties arises from its ability to detach from its core and to get spread within the whole nucleus. In doing so, the heavy core preserves its identity but the shell, which is relatively very light, looses it.

In effect, the low energy of the neutron shell makes it easily captured and shared by other nucleons, which consequently get bound together. Nuclear neutrons loose their identity and so they can be referred at the most as pseudo-neutrons. In fact the neutron only exists in the free state, and in such state it is unstable, i.e. its shell detaches from its proton core and undergoes a transition into a free orbital corresponding to the electron. Inversely, within the nucleus a spread shell may be recaptured by a proton, leading thus to a neutron.

### III.3. Hydrogen Atom

The hydrogen atom represents another type of proton-electron association. The H atom and the neutron may be seen as two different configurations of the same dual system $p^+, e^-$. In both cases, the proton acts as a core, but in the H atom the electron preserves its identity and wraps the proton at a very large distance of some $10^5$ times the proton size. In such a configuration the epicenters of the proton ($\Psi_p^+$) and the electron ($\Psi_e^-$) structuring orbitals are not concentric (the electron structure epicenter being peripheral), unlike the neutron ones (whose $\Psi_p^+$ and $\Psi_s^-$ are epicentral).

### III.4. Atomic Nucleus

The atomic nucleus can be seen as a cloud of anticorpuscles $c^+$ describing orbitals corresponding to the proton immersed itself in a cloud of corpuscles $c^-$ describing unspecific orbitals wrapping the agglomerate of protons and acting as bonding carriers. The unspecificity of the orbitals spun by the $c^-$ corpuscles means that these orbitals do not correspond to any specific particles but they instead wrap the nucleus. In other words, the $c^-$ corpuscle describes a collective orbital whose quantum state does not correspond to any free particle. In conventional terms of protons and neutrons, the nucleus can then be seen as formed by a core of protons and denatured neutrons, i.e. of neutrons having delivered their shell.

Nuclear neutrons are considered to lead to bare protons, which join the rest of nuclear protons, leading hence to a core exclusively formed by protons. The shells of the dissociated neutrons in getting spread within the core are shared by the core protons, tying them together. The ability of the neutron shell to dissolve within the nucleus and to be shared by the nucleons constitutes a ground stone of the orbital model. The nuclear forces are considered to be carried by the neutron shell by getting spread into the whole nucleus. In other words, nuclear strong interactions are generated by the $c^-$ orbitals wrapping the nucleus.

### III.5. Hydrogen $_1H^2$ and Helium $_2He^3$

These two nuclides are isotones, i.e. from a standard point of view they have the same neutron content, reduced to a single neutron. In the orbital context, since nuclear neutrons are considered to dissociate into their two constituents, core and shell, it thus means that both nuclides have homologue shells, generated by a single corpuscle, but differ in their core respectively composed of two and three protons. The shell with a single corpuscular carrier may have three different cores containing one, two, or three protons, and corresponding to the neutron, the deuteron, and the helion $_2He^3$. Consequently the wrapping orbital is differently stretched along with the different cores size, and therefore the resultant different degrees of stress confers them different net energies.

The deuteron $_1H^2$, conventionally considered to be composed of a proton and a neutron, is in the orbital context composed of a two proton core wrapped by the dissolved neutron shell which is then shared by the two protons (Fig.1). Since the mass of two protons weights 2.01455 amu and the deuteron has a mass of 2.01355 amu it has hence a mass defect of $-1.000 \times 10^{-3}$ amu which corresponds, on the part of the wrapping orbital, to a net negative energy of -0.93 MeV. The deuteron stability stands on the balance between two antagonist forces, the attractive electrostatic forces between the wrapping orbital and the core, and the repulsive electrostatic forces between the two protons of the core.

The helion $_2He^3$, being conventionally composed of one neutron and two protons, is thus in the orbital version composed of a three protons core and a single corpuscle wrapping orbital. The helion has a mass of 3.014933 amu and since the mass of three protons corresponds to 3.021830 amu, the wrapping orbital has thus a net energy of $-6.897 \times 10^{-3}$ amu or -6.42 MeV. Its stability stands on the balance of the attractive electrostatic forces between the wrapping orbital and the core and of the repulsive electrostatic forces between the three protons of the core. The effective stability of this single-corpuscle wrapping orbital manifests that it has enough energy to tie together the three core protons, i.e. the bonding energy ($E_1$) between the orbital and the core is greater than the dispersive energy ($E_2$) within the core:

**E = E$_1$ (core-shell bonding energy) - E$_2$ (core inner dispersive energy**

(in which $E_1 > E_2$ )                                     (4)

### III.6. Hydrogen $_1H^3$ and Helium $_2He^4$

These two isotones contain two neutrons, meaning that in the orbital frame they have homologue wrapping orbitals formed by two corpuscles. This dual orbital can confine from three up to eight protons. The triton $_1H^3$, composed of one proton and two neutrons, is equivalent in the orbital frame to a three-proton core and a double wrapping orbital proceeding from the two dislocated neutron shells. Although the shell wrapping the three protons is formed by two orbitals, such an association is nevertheless unstable and degrades. The triton instability derives from the balance, on one hand from the repulsion between the two orbital corpuscles and the repulsion between the core protons, and on the other hand, from the attraction between core and wrapping orbital. The shell net cohesive energy corresponds to the balance between two antagonist energies, one cohesive ($E_1$) and the other one dispersive ($E_2$), derived from two sets of opposite forces. The cohesive energy ($E_1$) raises from the attractive force between the positively charged proton core and the negatively charged shell. The dispersive energy ($E_2$) raises from two sources, the repulsive forces between the core protons ($E_{21}$) and the ones between the shell carriers ($E_{22}$). The shell net bonding energy is:

**E = E$_1$ (core-shell) - [E$_{21}$ (core) + E$_{22}$ (shell)]**            (5)

On the other hand, this net energy is equal to the nucleus mass defect:

**E = M∗c$^2$ - (m$_p$∗c$^2$)∗A**                               (6)

where M, $m_p$, A , and c are respectively the nucleus mass, the proton mass, the mass number, and the celerity of light. At first it may seem strange that the triton $_1H^3$ with its double orbital results unstable while the helion $_2He^3$ with a single orbital is stable, both having identical cores composed of three protons. However the explanation is straightforward and comes up from the electrostatic repulsion between the two corpuscles of the wrapping orbital, which weakens its effective bonding efficiency. It ends up that the net force is not strong enough to make this system stable.

The $_2He^4$ and the $_1H^3$ nuclei have a wrapping orbital containing two negative corpuscles, since both have two dissolved neutrons. However, although they have wrapping orbitals containing the same number of corpuscles, the $_2He^4$ is exceptionally stable in contrast to the unstable $_1H^3$, whereas in the latter case the wrapping orbital must only bind a three-proton core instead of a four-proton one. This difference rises from the balance between the repulsive forces within the core and within the wrapping orbital and the attractive forces between wrapping orbital and core. Let us point out another factor which affects the nucleus cohesive energy and thus the stability: a factor that instead has to do with the spin. In effect, since the $_2He^4$ proton core has four half-integer spins and its shell two integer ones, hence the system can easily match them to achieve a null net spin.

### III.7. Short Range Interactions: Strong and Weak

Short range interactions of both types, strong and weak, proceed from the orbital nature of elementary particles and are directly generated by their structuring orbital. A restricted analogy extracted from the atomic scale is provided by the covalent forces whose short range is limited by the atomic orbital size. The structure of the bonding shell of small nuclei is deeply submitted to quantum effects, while for large nuclei these effects are weakened and they rebound on nuclear interactions. The differences between the roughly typified weak and strong interactions would rise from the absence or presence of a shell and from its different quantum states, determined by its own characteristics (e.g. its carrier content and odd or even value) and also by its dependence on the proton core (e.g. its proton content).

An example of weak interaction is provided by the disintegration of the neutron, which in the orbital frame corresponds to the collapse of its shell, which is dependent on its own characteristics and on its relationship with the inner core. The neutron can be seen as a system composed of a core containing a sole proton and a shell held up by a single corpuscle, in a quantum state presenting an energy excess of 0.78 MeV, and thus can be considered an excited state with respect to the ground state constituted by the massive products of its disintegration, i.e. the free electron and proton.

An example of strong interaction is provided by the deuteron, whose cohesion is generated by the wrapping orbital of its two protons core. The range of the interaction is fixed by the extension of the deuteron single-carrier orbital and its strength is fixed by the core and the wrapping orbital self- characteristics, and also by their interdependence. The nucleus cohesion is determined by the balance of, the attractive forces between shell and core, and the dispersive forces inner to both. However for the particular case of the deuteron there is no repulsive force inner to its shell since it only contains a single carrier.

## IV. NEUTRON STRUCTURE, CHARGE DENSITY and CLASSICAL RADIUS

### IV.1. The neutron and proton radial charge density

Let us focus here on an experimental information concerned with the radial dependence of the charge density of the proton and the neutron, as shown in figure IV.1. Although it has been a text book knowledge for years (24), its deep relevance and significance have not been clearly understood and still less satisfactorily used. The Standard Model omission of this crucial experimental data is incomprehensible.

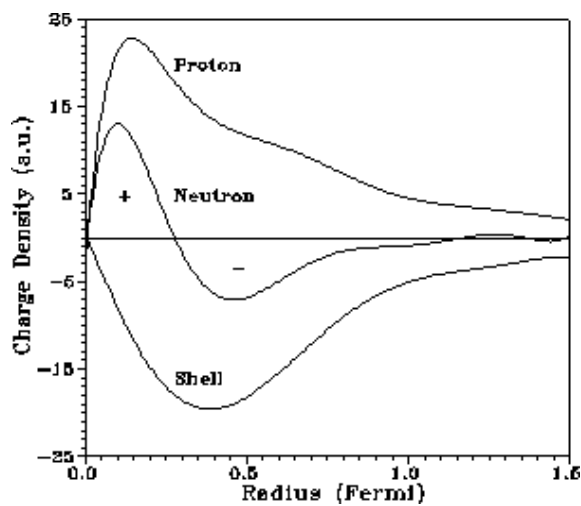

*Fig.IV.1: Radial dependence of the charge density of the proton and the neutron. The neutron clearly shows a positive core (zone +) and a negative shell (zone -). The radial charge distribution of the neutron shell has been deduced from the neutron and the proton charge distribution.*

It is well known that the nucleonic potential becomes repulsive at a distance smaller than a Fermi, evidencing hence the presence of a repulsive hard core. In the orbital conception of elementary particles it has been conferred to this experimental information all the relevance that it deserves. Within its context, the neutron is considered to be formed by a positive core, which is nothing else but a proton, and by a negative shell. The proton is itself structured by the orbital spun by a $c^+$ corpuscular carrier. The shell is instead structured by an orbital spun by a $c^-$ carrier. It is considered to be very reactive due to its affinity to be shared with other neighboring protons. For example, the neutron shell (unstable) has a strong trend to accommodate a second proton in its core, which increases its stability, leading to the deuteron (stable). In further building steps the primary trend is to preserve two protons for each $c^-$ shell carrier, which acts as the cohesive element of the nuclear structure. The departure from this primary trend is due to secondary effects, e.g. shielding and saturation.

The orbital conception of the neutron is in agreement with Figure IV.1 in which its radial distribution of the charge density clearly evidences a positive core (zone +) and a negative shell (zone -). The proton core is the cause of the repulsive behavior at a shorter radius than the Fermi and the shell plays instead the role of a bonding element. The charge distribution of the shell (added in the figure) has been directly deduced from the one of the proton and the neutron.

The quark model of the neutron (u,d,d) with fractional charges (+2/3, -1/3, -1/3) does not implicitly predict a positive core and a negative shell neither the nucleonic repulsive core potential, and is hardly able to account for it without appealing to highly artificial and twisted arguments, as always do the quark model and its supporting QCD, such as all their ad hoc properties, e.g. the quark fractional electric charges, their six flavors, the three colors of each favor, the height types of gluons, etc...

### IV.2. The neutron classical radius

The classical radius of the neutron can be derived straightforwardly from the potential energy. In effect, the orbital conception of the neutron assumes that it is made of a core and a shell. Since the core is considered to be a proton whose structural orbital is spun by a corpuscular carrier with positive electric charge ($c^+$), and since the shell is instead considered to be spun by a corpuscular carrier of negative electric charge ($c^-$), the negative shell has hence an electrostatic potential energy with respect to the positive core.

On another hand, it is known that the experimental mass difference between the neutron and the proton is equal to 1.29 MeV/$c^2$, which corresponds hence, in the orbital frame, to the mass of the shell or equivalently to a shell energy of 1.29 MeV. Let us now consider a corpuscle $c^-$ falling from infinity into the field of a proton, until acquiring a potential energy of 1.29 MeV. In a Coulomb field the potential energy E is equal to:

$$E = \int F(x)dx = \int (q^2/x^2)dx \qquad (7)$$

Hence, a unitary electric charge q falling in a Coulomb field from infinity down to a distance r to the field epicenter (e.g. a proton) will acquire a potential energy: $E = q^2/r$. Since the neutron shell has an energy of 1.29 MeV, thus:

$$r = q^2/E = 1.1 * 10^{-15} \text{ m} = 1.1 \text{ Fm} \qquad (8)$$

i.e. the classical radius of the neutron is equal to about 1.1 Fermi. The orbital approach to the neutron structure provides a simple way to get its classical radius, whose value is in good agreement with experimental data, in particular if compared to the one of 1.07 Fm of the mean electromagnetic radius of nucleons (25).

### V. NUCLEAR FUSION, RADIOACTIVITY $\beta^-$, $\beta^+$, $\gamma$ AND ELECTRON CAPTURE

### V.1 Nuclear Fusion from H, D and T

The fusion of a proton and a neutron goes by way of strong interaction, forming the deuteron. In the orbital context, this ensues from the neutron wrapping orbital or shell which gets shared and acts as a link. The underlying process is straightforward since it corresponds to the reaction of the neutron shell with the nearby proton, leading to a new wrapping shell common to both protons: the bare proton and the neutron core proton. The deuteron appears thus formed by a two-proton core wrapped together by the original neutron shell, which is thence shared by both. The deuteron may thus be regarded as an inflated neutron with a core of two protons instead of only one.

The fusion H,H of light hydrogen, i.e. of two protons into a deuteron is highly difficult and goes through weak interaction which confers on the process a very small cross-section. According to the orbital conception of elementary particles, since protons are constituted by a bare-core orbital, i.e. without any wrapper, they are hence unable to bond together without previously getting a bonding shell. This is the main reason that makes their fusion so uneasy, and now the question is how do they get a bonding shell in order to have a chance to bond during collision. To do so, part of their kinetic energies must be materialized into pairs of corpuscle-anticorpuscle $(c^+,c^-)$ when colliding. Then, if a corpuscle $(c^-)$ is captured by the two colliding protons it forms a common wrapping orbital which bonds them together.

Alternatively, if we consider that a corpuscle $(c^-)$ is at first captured by one of the two protons it turns into a neutron and thus at this step of the process it may be schematically assimilated to a fusion between a proton and a neutron. On its turn, the anticorpuscle $(c^+)$, formed conjointly to the corpuscle $(c^-)$, is rejected and its orbital structure acquires the $\beta^+$ identity. Besides, in seeking spin conservation, a neutrino is also emitted, and since the neutrino is here considered to be made of a pair corpuscle-anticorpuscle, it ends up that two corpuscle-anticorpuscle pairs must materialize during the proton's collision to be able to fuse. The limiting factor which fixes the fusion cross-section corresponds thus to the probability of the colliding protons to create two pairs $(c^+,c^-)$ and to capture one of the corpuscles $(c^-)$.

The H,D fusion differs from the H,H one mainly in the fact that there is already a wrapping $(c^-)$ orbital, the deuteron one. The fusion requires thus only the deuteron $(c^-)$ wrapper to be shared with the bare proton, leading so to the $_2He^3$ nuclide, formed by three core protons bonded by a single $(c^-)$ wrapping orbital. Besides, during the collision a pair corpuscle-anticorpuscle is formed and ejected in form of $\gamma$.

In the H,T fusion the triton $_1H^3$ provides a double wrapping orbital $(c^-,c^-)$. Thus, the fused system ends up formed by a four core protons sharing a bonding shell containing two $(c^-)$ corpuscular carriers, corresponding to the orbital scheme of the helium nucleus $_2He^4$.

In the D,D fusion each colliding nuclide has a wrapping orbital with one $(c^-)$, which hence leads to a double wrapping orbital $(c^-,c^-)$ with a four protons core, system which is stable and corresponds to the helium $_2He^4$.

In the D,T fusion the deuteron single $(c^-)$ wrapping orbital reacts with the triton double wrapping one. The fusion depends on the wrapping orbital's reactivity and the lifetime of the fused system depends on the acquired quantum state and on the allowed disintegration channels. The transitory system formed is made of five protons wrapped by three $(c^-)$ carriers, which turns out to be unstable and dissociates into a neutron and helium $_2He^4$ nucleus.

In the D, $_2He^3$ fusion the deuteron single $(c^-)$ wrapping orbital reacts with the also single $(c^-)$ wrapping orbital of the helion $_2He^3$, leading in a transient way to a double $(c^-)$ orbital wrapping a five-proton core, system which is unstable and disintegrates into helium $_2He^4$ through expulsion of one proton.

In the T, $_2He^3$ fusion the tritium provides a double $(c^-)$ wrapping orbital and the $_2He^3$ a single $(c^-)$ one. The orbitals' reaction transitorily leads to a triple $(c^-)$ wrapping orbital with a six-proton core. In this arrangement, the attractive forces between core and shell are not strong enough to overcome the repulsive ones inside the core and to the shell and thus the system vanishes instantaneously.

### V.2. Radioactivity $\beta^-$, $\beta^+$, $\gamma$ and Electron Capture

The radioactivity $\beta^-$ corresponds, from the orbital stand point, to the ejection of a carrier corpuscle $(c^-)$ from the nucleus wrapping orbital. Once free, the corpuscle acquires an intrinsic structuring orbital which corresponds to the electron. Of all nuclei, the neutron, which can be seen as a one nucleon nucleus, constitutes the simplest case of radioactivity $\beta^-$, corresponding to the disintegration of the neutron wrapping shell.

The electron capture corresponds to the inverse process, i.e. when an electron falls into the nucleus its body orbital dissolves and its corpuscular carrier $(c^-)$ acquires then an extrinsic orbital which runs on the nucleus, acting consequently as nucleons linking carrier. In terms of orbitals the process leads to the transition of the $(c^-)$ corpuscle from its intrinsic orbital, corresponding to the electron, to an enlarged one which extends within the whole nucleus.

The radioactivities $\beta^+$ and $\gamma$ have a unique origin corresponding to two manifestations of the same process. Both arise from the creation of a pair corpuscle-anticorpuscle ($c^-,c^+$) within the nucleus. In the radioactivity $\beta^+$, the pair dissociates and the anticorpuscle ($c^+$) is ejected in form of positron, while the corpuscle ($c^-$) remains within the nucleus and passes to widen the nuclide wrapping orbital. In the radioactivity $\gamma$ the pair corpuscle-anticorpuscle created remains bond together and is ejected in the configuration corresponding to the photon.

Let us stress that the radioactivity $\beta^-$, on one hand, and the radioactivities $\beta^+$ and $\gamma$, on the other hand, obey thus to different processes. While the radioactivity $\beta^-$ corresponds solely to the expulsion of a corpuscle ($c^-$) already present in the nucleus, the radioactivities $\beta^+$ and $\gamma$ imply the previous creation of a pair ($c^-,c^+$).

## VI. NATURE and SATURATION of NUCLEAR FORCES

The nature and saturation of nuclear forces are regarded from the perspective of the orbital context. It differs from the standard one which coarsely considers that neutrons preserve their identity within the nucleus. Instead, the orbital theory of elementary particles considers the nuclear neutron to dissociate into its two components, a proton core and a shell. So, it provides an original and subtle conception of the physical underlying grounds governing the periodical table of elements and brings new bases for the laws controlling the nucleus stability. It also fixes a limit to the nucleus size and thus a threshold to heavy nuclides, and it brings a straightforward understanding of the saturation of the nuclides' content within isotone, isobar, and isotope families. The saturation of nuclear forces splits up into two separate sources, leading to two separate saturations, i.e. the saturation of the proton content of the nucleus core and the saturation of the corpuscular carriers content of its wrapping and cohesive shell.

In regard to the shell cohesive energy it merges three main quantities of fundamental physical relevance, i.e. the total orbital energy ($E_T$), the orbital energy per nucleon ($E_A$) and the orbital energy per (conventional) neutron ($E_N$). The total orbital cohesive energy is defined as equal to:

$$E_T = ((m_p * A) - M) * c^2 \qquad (9)$$

where $m_p$, $M$, $A$, and $c$ represent the proton mass, the nucleus mass, the atomic number and the speed of light.

These three quantities ($E_T$, $E_A$, and $E_N$) acquire a deep specific significance, and are essential to apprehend the characteristics and behavior of nuclides. They respectively define the total cohesive energy ($E_T$) provided by the whole wrapping corpuscular carriers of the shell, the mean cohesive energy ($E_A$) perceived by each core proton, and the mean cohesive energy ($E_N$) that delivers each wrapping corpuscular carrier. Their behaviors, observed through their evolution within isotope, isotone, and isobar families, allows to profile the scheme of their specific characteristics. Let us thus focus on the evolution of these three basic orbital quantities within one type of isofamilies, the isotones, and check if their respective behaviors are consequent with the orbital stand point.

Let us recall that the net cohesive energy ($E_T$) of the wrapping orbital corresponds to the balance between two antagonistic energies ($E_1$ and $E_2$), derived from two opposite sets of forces.

$$E_T = E_1 - E_2 = E_1 - (E_{21} + E_{22}) \qquad (10)$$

The cohesive energy ($E_1$) results from the attractive forces between the positively charged protons core and the negatively charged wrapping orbital. The dispersive energy ($E_2$) raises from two sources: the repulsive forces between the core protons leading to a core dispersive energy ($E_{21}$) and the ones between the shell carriers leading to a shell dispersive energy ($E_{22}$). The cohesive energies $E_T$, $E_A$ and $E_N$ of the wrapping orbital represent the main parameters fixing the stability of the nucleus and respectively express the total net energy, the net energy per nucleon, and the net energy per neutron of the wrapping orbital. The reinterpretation of the nature of nuclear forces, based on the spreading of the neutron's shell all over the nucleus, has been checked over the totality of nuclides, which from reference (2) comprises 2226 nuclides.

Here, the sole case of isotones will be considered at following, yet isotopes and isobars have been reported elsewhere (29).

## VI.1. SATURATION OF NUCLEAR FORCES WITHIN ISOTONE FAMILIES

Isotones keep a fixed number of dissociated neutron, while the protons content varies and so does the nuclei net electric charge. Since the proton is barehanded, i.e. not provided with wrapping orbital or shell, its inclusion into the diverse nuclides of an isotone family maintains constant the number of wrapping corpuscular carriers. Nevertheless the wrapping orbital energy is affected, since a varying number of core protons rebound on the attractive forces between it and the core. In other words, while the negative charge of the shell remains constant, the positive charge of the core varies and affects thus the strength of the mutual core-shell attraction.

However, if an increment of protons within the core does increase the attractive strength between core and shell, it does not lead necessarily to an increment of the orbital bonding energy and in fact it presents counterparts leading to the weakening of nuclear forces in two opposite limits. Since within any isotone family the shell has a constant carrier content, one limit arises from a so low proton content that the attractive forces between core and shell end up being too weak to overcome the repulsive forces within the constant shell.

Schematically, it may be considered that the system disintegrates by explosion of the shell. Inversely, the other limit arises from a so high proton content that the repulsive forces within the core are excessive and overcome the attractive bonding forces between the core and the wrapping orbital, i.e. they overthrow the bonding capacity of the constant shell. In this second case it may be said that the system disintegration arises from the core explosion.

### VI.1.a. Saturation of the orbital total energy vs. proton content

Figures VI.1.1 and VI.1.2 evidence the raise of the nuclear total bonding energy ($E_T$) up to saturation along with the increment of the nucleus proton content, within isotones. In the orbital context this variable number of protons not corresponds to the part of core protons which are compensated by the corpuscles of the wrapping orbital. An increment of the proton content increases the electrostatic forces between

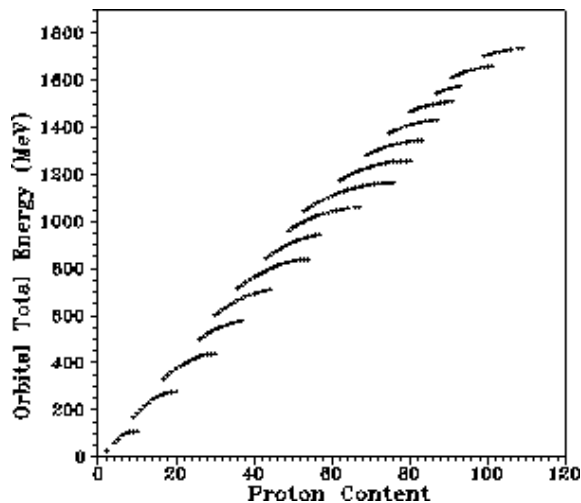

*Fig.VI.1.1:* Orbital total cohesive energy vs. proton content, for the isotones with neutron contents from 7 to 157 with step 10.

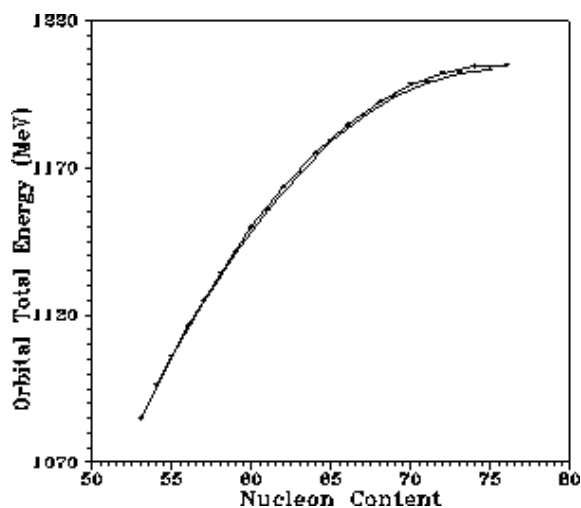

*Fig.VI.1.2:* Orbital total cohesive energy vs. proton content, for the isotones with 87 neutrons.

the core protons and the wrapping corpuscular carriers, however it simultaneously increases the repulsive forces within the core. The saturation of the shell total bonding energy occurs when the increment of the proton content does not improve any more the net orbital energy. In other words, the energy gain of the wrapping orbital saturates through the incorporation of protons and consequently the shell becomes unable to confine more protons within the core.

**VI.1.b. Orbital mean energy per nucleon vs. nucleon content**

Figures VI.1.3 and VI.1.4 stress the fact that the shell bonding energy per nucleon ($E_A$) presents a maximum, which favors maximum stability of the nucleus. In the orbital context, the shell bonding energy per nucleon derives from the mean presence density of the bonding carriers in the neighborhood of each core proton. In other words, the core protons are immersed into a cloud formed by the carriers' density of presence, which generates the wrapping shell.

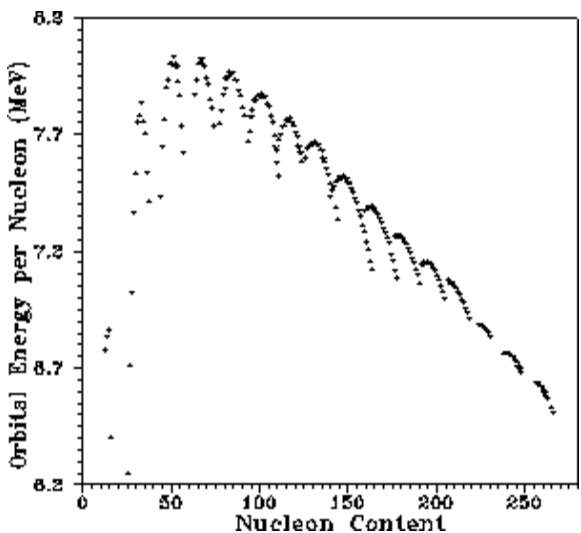

*Fig.VI.1.3:* *Orbital cohesive energy per nucleon vs. nucleon content, for the same isotones than in fig.VI.1.1.*

The carrier density presents an optimum which leads to a maximum bonding energy of the shell. When the proton content first raises so does the energy balance derived from antagonistic effects due on one hand to an increased repulsive forces within the engrossing proton core and on the other hand to the increased electrostatic field from the core which consequently intensifies the shell-bonding energy. After reaching a maximum the net energy balance decreases, stressing that for a still increasing density of presence of carriers the repulsive electrostatic forces among them raises in such a way that they prevail and saturate the net total bonding energy ($E_T$), consequently decreasing the energy per core proton ($E_A$) when the number of protons further increases.

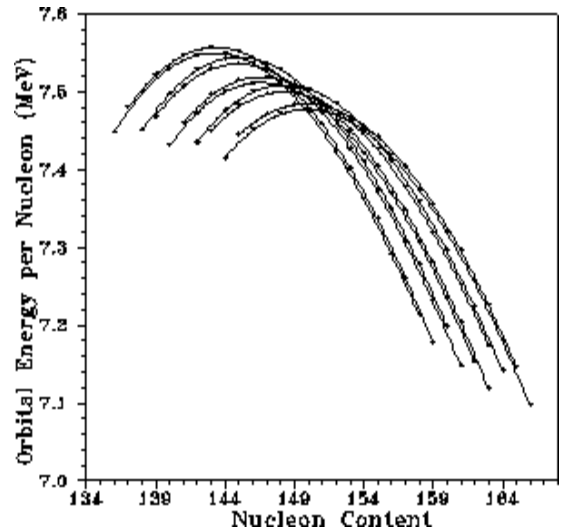

*Fig.VI.1.4:* *Orbital cohesive energy per nucleon vs. nucleon content, for the isotones with 85 to 89 neutrons.*

**VI.1.c. Orbital mean energy per neutron vs. proton content**

The mean orbital energy per neutron ($E_N$) represents in the orbital context a fundamental quantity since it corresponds to the mean bonding energy carried out by each shell carrier and which is the source of the nucleus stability. Figure VI.1.5 evidences that this quantity increases along with each proton added, showing some tendency to saturate. Saturation occurs when the mean cohesive energy of each carrier ($E_N$) becomes too weak to sustain the growing proton core, due to the increasing dispersive forces with the core.

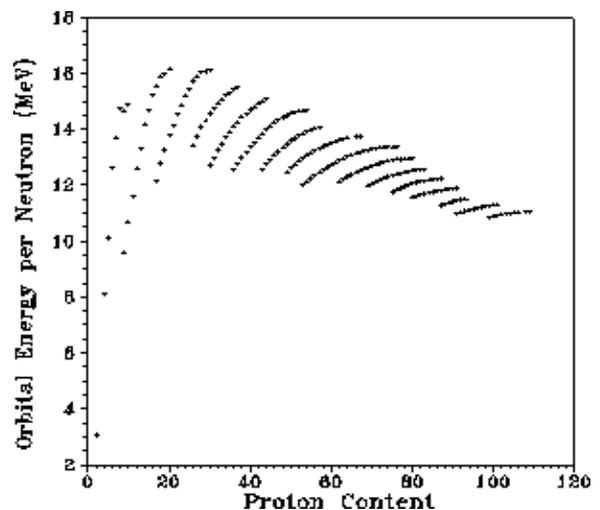

*Fig.VI.1.5:* *Orbital cohesive energy per neutron vs. proton content, for the same isotones than in fig.VI.1.1 and fig.VI.1.3.*

### VI.1.d. Splitting of the orbital energy vs. even or odd value of the proton content

For isotones this splitting effect is always present whatever the proton content is even or odd. Since for isotones only the core is affected through a varying proton content, the splitting of the shell cohesive energy is thus promoted by the spin interactions proceeding from the varying core. Figure VI.1.4 evidences the splitting of the orbital energy per nucleon ($E_A$) along with an even or odd number of protons. This split effect is observed in all isotone families, i.e. it is independent of the neutron content and of its even or odd value. The evolution of higher energies corresponds to an even number of protons and the one of lower energies to an odd number, whatever the even or odd value of the neutrons content of the isotone families. Since the carrier content of the shell is constant, this splitting is thus induced by the core, whose proton content represents the variable parameter.

## VII. COMMENTS

### VII.1. About the presumed experimental evidence of quarks

There is an enormous mathematical interpretation of quark-claimed evidence, but no direct observation. Quarks are bound particles within hadrons, so they would be virtual particles at the most. To acquire the status of real particles they should get free from their confinement and in order not to mistake them with other real particles their fractional charge should also be observed. A mathematical fitting with experimental results is not a sufficient requirement for an identification to physical reality. To illustrate this standpoint let us give the example of nuclear forces.

For a long period of time they have been formulated within the Yukawa interpretation of short range interactions. Good mathematical fittings with experiments were obtained in many cases. So, the associated physical interpretation of reality was an interchange of pions (and also of kaons). This physical interpretation, which has been considered for quite a while as the correct one, has now been abandoned. So, if pions which are real particles have fallen from grace as strong-interaction carriers after a long mathematical support, why quarks, which are not even real particles but only virtual ones, should be kept out of suspicion, taking also into account that their mathematical backing is highly complex, artificial and doubtful. To say nothing about the associated and necessary gluons, still more artificial particles. We think that the scientific community should be more critical and cautious about considering them as real particles (26). The standard model may end up representing an unfortunate example of theorists' strategy to approach physical reality, based on a mathematical hypertrophy which has lead to an atrophy of a previous settlement of reliable conceptual grounds.

### VII.2. About physical reality

Most theorists don't care to understand the physical reality underlying physical experiments, but instead they are just concerned with developing a mathematical formulation that fits with the experimental results. It will be always possible to mathematically fit a set of experimental results if the formulation is complex enough and contains multiple parameters (such as the 18 ones of the standard model). The higher is the complexity of the mathematical formulation the more distant it is from a realistic representation of physical reality. Moreover, the refutal of accessibility to physical reality or the disdain for its apprehension leads to an attitude that deeply impoverishes the aims of physics.

### VII.3. Antiparticles

Antiparticles are straightforwardly described since the $c^+$ and $c^-$ carriers structure both particles and antiparticles. For example, the electron (particle) is structured by a $c^-$ carrier and the positron (antiparticle) by a $c^+$ carrier. Instead the proton (particle) is structured by a $c^+$ carrier and the antiproton (antiparticle) by a $\bar{c}$ carrier. What differs an electron ($e^-$) from an antiproton ($p^-$) and a positron ($e^+$) from a proton ($p^+$) is the quantum state of their structuring orbital. Neutral particles and antiparticles, both formed by a pair ($c^+,c^-$), differ in that the orbital of the $c^+$ and $\bar{c}$ carriers can be in different quantum states which are inverted, e.g. $[c^+(|1>),c^-(|2>)]$ for a particle and $[c^+(|2>),c^-(|1>)]$ for the corresponding antiparticle.

Physics Essays      Vol.12, no.2, 1999

### VII.4. About the electron, proton and neutron

The method used to calculate the neutron radius (more specifically the one of its shell) cannot be used for the electron or the proton whose radius is differently derived (1). The reason stands in that the neutron is considered to have a proton core and thus the standard electrostatic formulation can be used.

This is not the case for the electron and the proton which have no core. They are structured by a single orbital which is considered to be differently sustained (1). Their structuring orbital is confined through the equilibrium of two antagonistic forces, one centripetal (a Lorentz like force) and a centrifugal force (restoring like force). These two forces are seen as action and reaction. Still, further quantitative developments (1) have lead to the unification of the electron, muon, proton, neutron and H atom, achieved through a magnitude $Q = m \mu r$ which ends up to have the same value for all them and equal to $Q = (e\, h\, r_o)/2$, (e and $r_o$ are the electron charge and classical radius and h the Planck constant (29).

### VII.4.1. Neutron creation

Let us consider a few reactions leading to neutron creation and interpret them from the orbital structure standpoint:

$$\nu + p \rightarrow n + e^+ \qquad (1)$$
$$\nu + p \rightarrow n + \mu^+ \qquad (2)$$
$$\gamma + p \rightarrow n + \pi^+ \qquad (3)$$

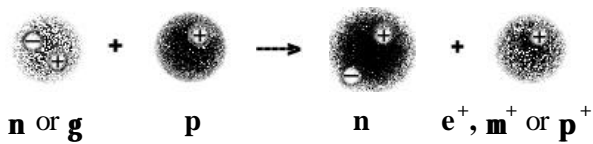

$\nu$ or $\gamma$    p    n    $e^+, \mu^+$ or $\pi^+$

These three reactions belong to the same archetype, i.e. a quantum ($\nu$ or $\gamma$) impinges on a proton and reacts with it, generating a neutron and a positively charged particle. Since all neutral particles are considered formed by two oppositively charged structural carriers and charged particles are considered dissociated neutral particles, thus structured by the orbital of a single carrier, hence the three positively charged particles $e^+$, $\mu^+$ and $\pi^+$ have the same structure ($c^+$) but are in three different quantum states. The neutron is here considered formed by a core proton structured by the positive carrier and a shell structured by the negative carrier.

In the $\nu$, p and $\gamma$, p reactions, the negative carrier $c^-$ of the quantum $|\psi(c^+,c^-)\rangle$, in form of $\nu$ or $\gamma$, is transferred to the proton, leading to a neutron and the $c^+$ carrier acquires diverse quantum states $|\psi(c^+)\rangle$, i.e. different structural orbitals $|\psi(e^+)\rangle$, $|\psi(\mu^+)\rangle$ or $|\psi(\pi^+)\rangle$, corresponding to $e^+$, $\mu^+$ or $\pi^+$.

Let us now interpret the following reaction:

$$e^- + p^+ \rightarrow n + \nu \qquad (4)$$

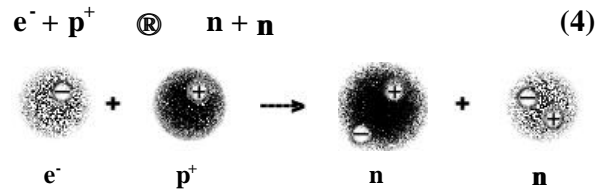

$e^-$    $p^+$    n    $\nu$

Here the impinging quantum is constituted by an electron. The reaction is quite similar to the previous ones but in this new case the incident particle contains only one structural carrier, which is transferred to the proton, thus leading again to a neutron. The excess energy of the reaction is expulsed by means of a quantum $|\psi(c^+,c^-)\rangle$ in the neutrino state.

Let us interpret the following reaction leading to the deuteron lease:

$$\gamma + d \rightarrow n + p$$

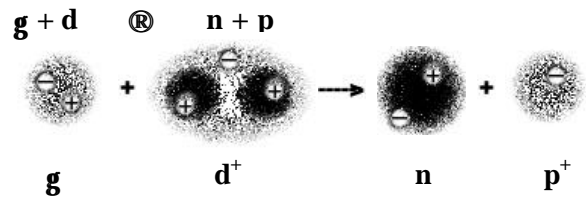

$\gamma$    $d^+$    n    $p^+$

Here the impinging quantum $|\psi(c^+,c^-)\rangle$ is a photon which is absorbed and releases its energy to the deuteron. In the induced lease of the deuteron $|\psi(c^+,c^+,c^-)\rangle$, one of its core proton $|\psi(c^+)\rangle$ gets free while the other one $|\psi(c^+)\rangle$ keeps the shell $|\psi(c^-)\rangle$, leading thus to a neutron $|\psi(c^+,c^-)\rangle$.

### VII..2. Neutron disintegration

$$n \rightarrow p^+ + e^- + \nu$$

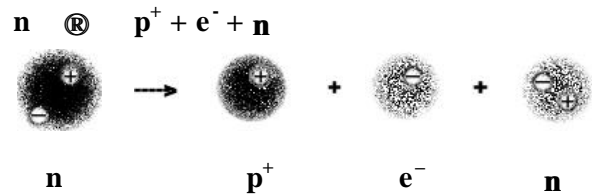

n    $p^+$    $e^-$    $\nu$

The free neutron is unstable and its disintegration is exothermic. The neutron looses its shell (of 1,29 MeV) which, in getting free, restructures into an electron (of 0,51 MeV) and the energy excess is materialized into a neutrino. The bare neutron core leads to a proton.

### VII.5. Paired (c⁺,c⁻) system and isolated (c⁺) and (c⁻) systems

The creation of any elemental particle is here considered to always go through the hop of a dual system formed by a pair of corpuscle (c⁺,c⁻) from a virtual to an energetic state, leading to a neutral particle. Space can be thought as populated of virtual quanta formed by self-existent corpuscle-anticorpuscle pairs in an aenergetic state. When energy is transferred to the virtual dual systems it causes their hopping to energetic states. During its transfer to the energetic state the dual (c⁺,c⁻) system may dissociate to form two separate single corpuscle systems (c⁺) and (c⁻), leading to the diverse charged particles according to the quantum state of their orbital structure (Fig. VII.1).

| Neutral Particle | Positive Particle | Negative Particle |
|---|---|---|
| $Y^\pm$ | $Y^+$ | $Y^-$ |

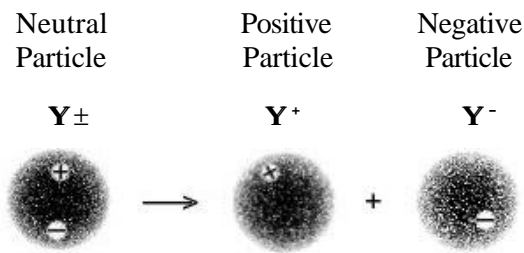

*Fig. VII.1*: *Partition of any neutral particle into two oppositely charged ones*

### VII.6. Evolution of the conception of nuclear forces

Nuclear forces have been at first conceived through the Yukawa interpretation and the cause of their short range was attributed to the mass of the carrier. So, at the discovery of the muon it has been considered to be the formulated massive carrier, but soon it has been disregarded and at the discovery of the neutral and charged pions these have been retained as the ascertained carriers. However at the discovery of the four kaons these have also been involved conjointly with the pions as carriers. Later with the rise of the Standard Model they all fell from grace in favor of the gluons, that auxiliary would also be the carriers of the inter-nucleons bonding forces.

What a poorly convincing evolution! The Yukawa interpretation of the cause of short range has ended up to be wrong for the strong interactions and we consider that it may also be inappropriate for the weak interactions. The Z and W particles may not be imperatively related to these interactions, as the muon, pions and kaons in regard to the strong ones.

Instead we have proposed that nuclear forces, strong and weak, both of short range, derive from the involvement of a shell. For instance, the neutron disintegration goes through the weak interaction because it corresponds to the loss of its shell (c⁻) which restructures in getting free, emitting then a quantum in form of an antineutrino. On its part, the p,p fusion into D goes through the weak interaction because protons' lack of a shell and their fusion requires the previous acquisition of one, which implies the creation of a pair of quanta (c⁺, c⁻) in form of a neutrino and an antineutrino, with the dissociation of one quantum (c⁺, c⁻) into (c⁺) and (c⁻) and the capture of (c⁻) which acts then as a shell. Instead the p,n fusion into D goes through the strong interaction because the neutron has already a shell ready to be shared, without the need of the previous creation and dissociation of any (c⁺, c⁻) quantum.

### VII.7. Conceptual unity of matter

The concepts developed upon matter fundamental blocks lead to a unitary conception of elementary particles. All particles have been conceived from a unique corpuscle, which manifests itself in two complementary forms, i.e. corpuscle and anticorpuscle. All particles and antiparticles derive directly from the type of corpuscle (c⁺ or/and c⁻) which generates them through its orbital. The particles formed by the orbital of a single corpuscle are thus electrically charged and those formed by a pair corpuscle-anticorpuscle are hence neutral.

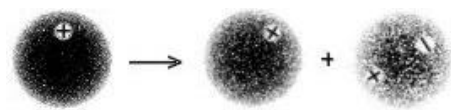

*Fig. VII.2*: *Particle decay to a lower mass one with emission of a neutral quantum*

The diversity of elementary particles is considered to derive from the multiplicity of quantum states of their structuring orbital, spun by the fundamental corpuscle. The particle rest energy is defined by the net energy of its structuring orbital. In any orbital exoenergetic transition the corpuscle emits a pair ($c^+,c^-$), therefore reproducing itself and its anti-itself (Fig. VII.2). This standpoint provides an extreme conceptual simplicity to the nature of elementary particles and to their genesis.

**VII.8. Asymmetrical composition of the Universe**

The Universe is observed to be asymmetrical with respect to its composition at the level of elementary particles. In effect, protons are not compensated by antiprotons and neither electrons by anti-electrons (positrons), besides the fact that their coexistence is incompatible. Within the standard conception of antimatter the diverse alternatives are: the Universe may have never contained antimatter, or for some unknown reason matter and antimatter may have been separated, or still the amount of matter may have overcome the one of antimatter and the actual material Universe would be the residue of their annihilation. The orbital conception provides a novel possibility. The duality matter-antimatter would be satisfied at the corpuscle level, i.e. the Universe would have been composed of an equal amount of corpuscles and anticorpuscles.

However the Universe would intrinsically possess an asymmetrical behavior at the orbital level. The positive anticorpuscle would have acquired a structural orbital corresponding to the proton while the negative corpuscle would have acquired the electron one. This asymmetrical behavior might be typified at the level of proton and antiproton by assuming a weak asymmetry between both particles, proceeding from their structuring orbitals and leading to opposite magnetic moments for the same spin orientation (Fig. VII.3). Such an intrinsic asymmetry would have, at a cosmological time-scale, lead to the mutation of antiprotons into fossil electrons.

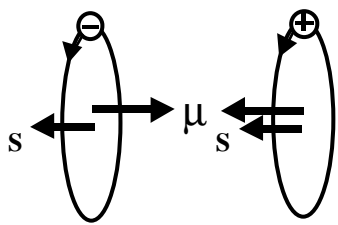

*Fig. VII.3*: *Partial asymmetry between particle and antiparticles (S-**m** asymmetry)*

For an identical spin orientation (i.e. identical giratory direction) negative and positive charges generate magnetic moments of opposite orientation. Since particles and antiparticles have opposite charges thus they have also opposite magnetic moments (with respect to their spin). This intrinsic S-$\mu$ asymmetry may or may not induce, depending on the specific structural quantum state of each type of particle-antiparticles, a weak difference of stability (presumably related to the CP violation).

The proton and the electron would thus be the only two stable massive end products of an intrinsic weak asymmetry between particles and antiparticles, that would have lead to a fundamental self-asymmetry of the Universe. This asymmetrical behavior would have been the Universe's solution to avoid self-annihilation. Although the structural orbital asymmetry between proton and electron still retains some basic mystery, at least symmetry between matter and anti-matter is preserved, being moved to a more fundamental level constituted by the duality corpuscle-anticorpuscle.

Whatever the underlying reason, we are confronted with the experimental evidence of a basic compositional asymmetry of the Universe, exclusively composed of two stable massive particles of opposite electric charge, the proton and the electron. Within the conceptual frame proposed, the fundamental asymmetry of the Universe still emerges through another of its facets, provided this time by the neutron. This third particle involved in the construction of matter, still preserves the asymmetry between proton and electron since the neutron is considered to be formed by the superposition of the proton and the electron structuring orbitals.

In such association the lighter electron orbital is affected by the heavier proton orbital, which implies that the electron has in fact lost its identity of free electron to turn into a new slightly heavier orbital wrapping the proton. Such new orbital with a mass of 1.29 MeV/$c^2$ forms then the neutron shell while the proton constitutes the neutron core. In the free state this association appears to be unstable since the free neutron degrades into a massive tandem proton-electron, but it is the master piece of nuclear structure by sharing its shell with protons.

### VII.9. Nature of free space or vacuum

We have considered up to now corpuscles and anticorpuscles as being the fundamental elements of matter and their intrinsic orbital as structuring elementary particles. The basic unit is formed by a corpuscle-anticorpuscle tandem constituting a neutral elementary particle, a unit which may get broken into its two components, leading then to a pair of charged elementary particles. Let us consider the dual system pairs $(c^+,c^-)$ and focus on its net energy. It has been expressed before that the net energy is the residue between the compressive (centripetal) and expansive (centrifugal) components of the orbital energy. If these do not perfectly match they lead to an energy residue which represents the net energy of the corresponding elementary particle.

However, it should not be discarded that the two antagonistic energy components of the dual system $(c^+,c^-)$ can fully compensate and therefore its net energy would then be null. Still more, diverse orbital quantum states might lead to a net null energy. Hence free space could be populated of quanta formed by virtual pairs $(c^+,c^-)$ of null net energy, which could eventuality constitute diverse populations with different aenergetic structural quantum states. Particles may then be regarded as proceeding from vacuum quanta which would have acquired energy through structural disequilibrium.

### VIII. CONCLUSION

Let us review some of the most relevant features of the unitary orbital theory and contrast a few of them with the quark equivalencies:

* The presented alternative orbital conception of elementary particles leads to a unitary conception of them and of their diverse interactions. It is based on a single fundamental corpuscle with a dual nature $c^+$ and $c^-$.

* The particle's orbital structure, embodied by the fundamental corpuscle, is an intrinsic characteristic and the corpuscle cannot be free of it. Each orbital quantum state defines a specific elementary particle and all its properties.

* Furthermore, this approach also leads to the conceptual unification of the four interactions. The two short-range interactions derive from the close contact of the particles orbital structure. The two long range interactions arise from the exchange of $(c^+,c^-)$ pairs in form of virtual photons and gravitons.

In order to check the foundations of the orbital conception of elementary particles and to contrast it with the quark one, let us point out some of its most specific implications.

* The first one is concerned with the structure of the proton which should evidence a single corpuscle of integer electric charge (however, in deep scattering experiments the creation of internal quanta $(c^+,c^-)$ may shelder the non composite structure of the proton). In contrast, from the Standard Model deep scatttering should evidence collision to fractional electric charges and of different sign, from the three quarks of the proton.

* A more specific and conclusive differentiation could emerge from the neutron. In effect, from its orbital conception the neutron is implicitly considered as formed by a positive core and a negatively charged shell. Hence its inner distribution ($\Psi_p^+$ and $\Psi_s^-$) of electric charges should evidence a splitting into these two oppositely charged orbitals. Instead its quark structure should manifest, such as for the proton, three fractional electric charges whose distribution should not fundamentally differ from the proton one and from which the subdivision into core and shell is not implied.

* Still another way to track back the two conceptions could stand on collision experiments on the deuteron. Its orbital conception predicts two positively charged cores being nothing else but two protons and a shell wrapping both of them, while the quark structure should manifest six punctual objects confined into two cores.

* An important application of the orbital conception applied to the neutron stands on the assumption of its dual structure, composed of a core and a shell. This simple hypothesis leads to a novel and straightforward qualitative understanding of nuclear forces and the nucleus structure, cohesion and stability.

* The neutron shell is unstable in the free state and its low energy of only 1.29 MeV makes it quite weak. However, it is considered to be very reactive, having a high trend to incorporate a second proton in its core. In this state the shell

becomes quite strong and stable. It may incorporate a third proton leading to the helion $_2H^3$ which is also stable. Thus, a shell composed by a single corpuscular carrier can confine from one to three protons, leading to the three isotones: the neutron, the deuteron and the helion $_2H^3$.

\* To built up heavier nuclei the shell must incorporate more than one carrier. So, a shell with two carriers can confine up to seven protons. The progressive increase of the shell carrier content allows a growing proton core. However the process presents a threshold which succeeds when the repulsive forces inner to the core or to the shell finally dominate over the attractive forces between them. At present the heaviest nuclide obtained (Z=111 and A=272) has thus, in the orbital frame, a core with 272 protons bonded by a shell with 161 carriers.

\* The presence of neutrons in the nucleus is indispensable since it constitutes the sole source of nuclear forces through the delivery of its shell. The orbital theory applied to nuclides provides a straightforward understanding of their behavior as isotones, isobars, and isotopes. Such a simple assumption as to confer a shell to the neutron allows to reinterpret the complete nuclear field.

\* A development (in accordance with quantum field theory, in particular with the fundamentals of quantum electrodynamics) has been proposed in which the concept of virtual neutral particles with null net energy made by ($c^+$,$c^-$) pairs is introduced as well as the hypothesis that they form a main background leading to a virtual Universe corresponding to free space. Elementary particles would represent energetic states extracted from the virtual ones and they would form a secondary foreground corresponding to the manifested Universe.

\* The composition asymmetry of the Universe, which is solely made of matter (i.e. without antimatter), takes a novel look through the orbital conception by establishing a symmetry at the corpuscular level and by introducing the asymmetry at the level of the particles structural orbital.

Finally, to sum up let us stress that the Orbital Conception of Elementary Particles leads to very concrete predictions in regard to their structure and interactions, which could be experimentally checked.

**(1)** The electron is structured by a single corpuscular carrier $c^-$ with a negative unitary charge.

**(2)** The proton is structured by a single corpuscular carrier $c^+$ with positive unitary charge.

**(3)** The neutron is predicted to have a composite structure which can be regarded as generated by the concentric superposition of an electron and a proton, leading to a new structure, comprising a core and a shell. The core is positively charged and constituted by the proton structural orbital (938,26 MeV) which is preserved. The shell is formed by a negatively charged orbital with an energy of +1,29 MeV, deriving from the electron structuring orbital (0.51 MeV) which is not preserved due to its low energy.

**(4)** The Deuteron structure is formed by two protons and a wrapping cohesive shell, negatively charged with an energy of -1,00 MeV.

**(5)** The Helion 3 is structured by a three proton core and a shell, negatively charged and with an energy of -6,42 MeV.

**(6)** The Triton is structured by a three proton core and a shell doubly negatively charged and with an energy of -5,89 MeV.

**(7)** As a generalization, the nuclear neutron is predicted to be dissociated into its two structural components and consequently the nuclear structure to be solely composed of protons bound together by a shell multi-negatively charged, whose cohesive energy derives from the difference between the mass of the nucleus and the mass of its proton core.

**(8)** The proposed photon structure is formed by two oppositely charged structural carriers and can be broken into its components, which in the free state restructure as positron and electron (Let us remind that the Standard Model has neither a structure for the photon nor for leptons). Let us quote novel experimental recent results about the rupture of (real) photons into a positron and an electron (27), data which are in perfect predictive agreement with the proposed structure of the photon ($c^+$,$c^-$) whose components take the positron ($c^+$) and the electron ($c^-$) structural state when free.

Since the proposed structure of the (real) photon is previous to these experimental results it can be regarded as a strong support to the prediction made about its structure. Furthermore, let us stress that the production of a $e^+,e^-$ pair in the collision of two real photons was first considered by Breit and Wheeler (28) who calculated the reaction cross section to be of the order of $r_o^2$, where $r_o$ is the electron classical radius. This value is the one attributed to the classical radius of the photon, electron and positron in their orbital conception, considered to be the larger radius an elementary particle can have. The fact that the reaction cross section depends precisely on $r_o$ brings another strong support to the photon and electron proposed orbital structure.

Predictions (3) and (7) have already got a remarkable backing from the experimental data of references (24) and (27). Confrontation with experiments is currently in development (29).

(*) E-mail: gsardin@intercom.es